\documentclass[12pt]{article}
\usepackage{pdproc,epsfig} 

\def\nn{\nonumber}
\def\bea{\begin{eqnarray}}
\def\eea{\end{eqnarray}}
\def\beq{\begin{equation}}
\def\eeq{\end{equation}}
\def\bq{\begin{quote}}
\def\eq{\end{quote}}
\def\gappeq{\mathrel{\rlap {\raise.5ex\hbox{$>$}} {\lower.5ex\hbox{$\sim$}}}}
\def\lappeq{\mathrel{\rlap{\raise.5ex\hbox{$<$}} {\lower.5ex\hbox{$\sim$}}}}

\def\NP{{\it Nucl.~Phys.~}}

\def\be{\begin{equation}}
\def\ee{\end{equation}}
\def\bc{\begin{center}}
\def\ec{\end{center}}
\def\bea{\begin{eqnarray}}
\def\eea{\end{eqnarray}}

\def\nn{\nonumber}
\def\gappeq{\mathrel{\rlap {\raise.5ex\hbox{$>$}} {\lower.5ex\hbox{$\sim$}}}}
\def\lappeq{\mathrel{\rlap{\raise.5ex\hbox{$<$}} {\lower.5ex\hbox{$\sim$}}}}
\begin{document}

\pagestyle{empty}
\begin{flushright}
{CERN-PH-TH/2004-076}\\
\end{flushright}
\vspace*{15mm}
\begin{center}
{\bf PROGRESS ON MODELS OF NEUTRINO MASSES AND MIXINGS}\\ 
\vspace*{1cm}
{\bf Guido Altarelli}\\
\vspace{0.3cm}
CERN, Department of Physics, Theory Division\\
CH - 1211 Geneva 23 \\
\vspace*{3cm}
{\bf ABSTRACT} \\ \end{center}
\vspace*{10mm}
\noindent
We review some recent results obtained by our group on models of neutrino masses and mixings in the general context of SUSY GUT's with an associated flavour symmetry
 \vspace*{5cm}
\noindent

\noindent
\begin{center}
{\it Talk given at the Fujihara Seminar on\\
Neutrino Mass and Seesaw Mechanism, 
KEK (Japan), 23--25 February 2004 }
\vspace*{2.5cm}
\end{center}
\begin{flushleft} CERN-PH-TH/2004-076 \\
May 2004
\end{flushleft}
\vfill\eject

\setcounter{page}{1}
\pagestyle{plain}



  



   
\normalsize\baselineskip=15pt

\section{Introduction}

In this presentation I will describe some recent results obtained in collaboration with Ferruccio Feruglio and Isabella Masina. In the first part, based on ref. \cite{afm03} recently updated including all available data, we discuss a quantitative study of the ability of models with different levels of hierarchy to reproduce the observed pattern of neutrino masses and mixings. As a flexible testing ground we
consider models based on SU(5)$\times$U(1)$_{\rm F}$. In this context, we have made statistical simulations of models with
different patterns from anarchy to various types of hierachy: normal hierarchical models with and without automatic
suppression of the 23 (sub)determinant and inverse hierarchy models.  We find that the hierarchical models have a significantly better success rate than those based on anarchy. The
normal hierachy models appear to maintain a considerable edge over inverse hierarchy or anarchy. 

In the second part, based on ref. \cite{afm04}, we discuss to which extent the observed mixing can arise from the diagonalisation of the charged lepton matrix. The neutrino mixing matrix $U$ is in general of the form $U=U_e^\dagger U_{\nu}$, where $U_e$ arises from the
diagonalization of charged leptons and $U_{\nu}$ is from the neutrino sector. We consider the possibility that $U_{\nu}$ is
nearly diagonal (in the lagrangian basis) and the observed mixing arises with good accuracy from $U_e$. We find that the fact that, 
in addition
to the nearly maximal atmospheric mixing angle $\theta_{23}$, the solar angle $\theta_{12}$ is definitely also large while at
the same time the third mixing angle $\theta_{13}$ is small, makes the construction of a natural model of this sort considerably 
more complicated. We present an example of a natural model of this class. 
We also find that the case that $U_\nu$ is exactly of the bimixing type is severely constrained 
by the bound on $\theta_{13}$ but not excluded. 
We show that planned experimental searches for $\theta_{13}$ could have a strong impact on bimixing models.

\section{Hierarchy versus Anarchy}

The smallness of neutrino masses interpreted via the see-saw mechanism \cite{seesaw} directly leads to a scale 
$\Lambda$ 
for L non-conservation which is remarkably close to $M_{GUT}$. Thus neutrino masses and mixings should find a
natural context in a GUT treatment of all fermion masses. The hierarchical pattern of quark and lepton masses, within
a generation and across generations, requires some dynamical suppression mechanism that acts differently on the
various particles. 
This hierarchy can be generated by a number of operators of different dimensions suppressed by inverse
powers of the cut-off $\Lambda_c$ of the theory. The different powers of $1/\Lambda_c$
may correspond to different orders in some symmetry breaking parameter $v_f$ arising from the spontaneous 
breaking of a flavour symmetry.
Here we describe some simplest models based
on SU(5) $\times$ U(1)$_{\rm F}$  which illustrate these possibilities \cite{u1}.
It is notoriously difficult to turn these models into fully realistic theories, due to
well-known problems such as the doublet-triplet splitting, the proton lifetime, the gauge coupling unification 
beyond leading order and the wrong mass relations for charged fermions of the first two  generations.
Some of these problems can be solved by adopting the elegant idea of GUT's in extra dimensions \cite{EDGUTs}. Here we adopt the GUT framework simply as a convenient testing ground for different neutrino mass scenarios.
 
\subsection{Models Based on Horizontal Abelian Charges}
 
We discuss here some explicit examples of grand unified models in the framework of a unified SUSY
SU(5) theory with an additional 
U(1)$_{\rm F}$ flavour symmetry \cite{afm03} . The SU(5) generators  act ``vertically'' inside one generation, while the
U(1)$_{\rm F}$ charges are different ``horizontally'' from one generation to the other. If, for a
given interaction vertex, the
U(1)$_{\rm F}$ charges do not add to zero, the vertex is forbidden in the symmetric limit. But the symmetry is spontaneously broken by the VEV's
$v_f$ of a number of ``flavon'' fields with non-vanishing charge. Then a forbidden coupling is rescued but is
suppressed by powers of the small parameters $v_f/\Lambda_c$ with the exponents larger for larger charge 
mismatch. We expect $M_{GUT} \lappeq v_f \lappeq \Lambda_c \lappeq M_{Pl}$. 
Here we discuss some aspects of the description of fermion masses in this framework. 

In these models the known generations of quarks and leptons are contained in triplets
$\Psi^{10}_i$ and
$\Psi^{\bar 5}_i$, $(i=1,2,3)$ corresponding to the 3 generations, transforming as $10$ and ${\bar 5}$ of SU(5),
respectively. Three more
SU(5) singlets
$\Psi^1_i$ describe the RH neutrinos. In SUSY models we have two Higgs multiplets $H_u$ and $H_d$, which transform as 5
and $\bar 5$ in the minimal model. The two Higgs multiplets may have the same or different charges.
In all the models that we discuss the large atmospheric mixing angle
is described by assigning equal flavour charge to muon and tau neutrinos and their weak SU(2) partners (all belonging
to the 
${\bar 5}\equiv(l,d^c)$ representation of SU(5)). Instead, the solar neutrino oscillations can be obtained with
different, inequivalent charge assignments.
There are many variants of these models: fermion charges can all be non-negative with only negatively charged
flavons, or there can be fermion charges of different signs with either flavons of both charges or only flavons of
one charge. We can have that only the top quark mass is allowed in the symmetric limit, or that also other third
generation fermion masses are allowed. The Higgs charges can be equal, in particular both vanishing or can be
different. We can arrange that all the structure is in charged fermion masses while neutrinos are anarchical.  

\subsubsection{F(fermions)$\ge$ 0}

Consider, for example, a simple model with all charges of matter fields being non-negative and containing one
single flavon ${\bar\theta}$ of charge F$=-1$. For a maximum of simplicity we also assume that all the  third generation masses
are directly allowed in the symmetric limit. This is realized by taking vanishing charges for the Higgses and for
the third generation components $\Psi^{10}_3$, $\Psi^{\bar 5}_3$ and $\Psi^1_3$.  
If we define F$(\Psi^R_i)\equiv q^R_i$ $(R=10,{\bar 5},1;~i=1,2,3)$,
then the generic mass matrix $m$ has the form
\beq 
m~=~
\left(
\matrix{ y_{11}\lambda^{q^R_1+q^{R'}_1}&y_{12}\lambda^{q^R_1+q^{R'}_2}&y_{13}\lambda^{q^R_1+q^{R'}_3}\cr
y_{21}\lambda^{q^R_2+q^{R'}_1}&y_{22}\lambda^{q^R_2+q^{R'}_2}&y_{23}\lambda^{q^R_2+q^{R'}_3}\cr
y_{31}\lambda^{q^R_3+q^{R'}_1}&y_{32}\lambda^{q^R_3+q^{R'}_2}&y_{33}\lambda^{q^R_3+q^{R'}_3}}
\right) v~~~,
\label{m1}
\eeq 
where all the $y_{ij}$ are dimensionless complex coefficients of order one and
$m_u$, $m_d=m_l^T$, $m_D$ and $M$ arise by choosing $(R,R')=(10,10)$, $(\bar{5},10)$,  
$(1,\bar{5})$ and $(1,1)$, respectively. 
We have $\lambda\equiv\langle{\bar \theta}\rangle/\Lambda_c$ and the 
quantity $v$ represents the appropriate VEV or mass parameter.
The models with all non-negative charges and one single flavon have particularly simple factorization properties.
For instance in the see-saw expression for $m_{\nu}=m_D^T M^{-1} m_D$ the dependence
on the $q^{1}_i$ charges drops out and only that from $q^{\bar5}_i$ remains.
In addition, for the neutrino mixing matrix $U_{ij}$, which is determined by $m_\nu$ in
the basis where the charged leptons are diagonal, one can prove that 
$U_{ij}\approx \lambda^{|q^{\bar 5}_i-q^{\bar 5}_j|}$, in terms of the differences of the
$\bar5$ charges, when terms that are down by powers of
the small parameter $\lambda$ are neglected. Similarly the CKM matrix elements are approximately 
determined by only the 10 charges \cite{u1}: $V^{CKM}_{ij}\approx\lambda^{|q^{10}_i-q^{10}_j|}$.
If the symmetry breaking parameter $\lambda$ is numerically close to the Cabibbo angle,
we can choose:
\beq
(q^{10}_1,q^{10}_2,q^{10}_3)=(3,2,0)~~~,
\label{cha10}
\eeq     
thus reproducing $V_{us}\sim\lambda$, $V_{cb}\sim\lambda^2$ and
$V_{ub}\sim\lambda^3$. The same $q^{10}_i$ charges also fix $m_u:m_c:m_t\sim \lambda^6:\lambda^4:1$. The
experimental value of $m_u$ (the relevant mass values are those at the GUT scale: $m=m(M_{GUT})$ \cite{koide}) 
would rather prefer $q^{10}_1=4$. Taking into account this indication and the presence of the unknown coefficients 
$y_{ij}\sim O(1)$ it is difficult to decide between $q^{10}_1=3$ or $4$ and both are acceptable. 
Of course the charges $(q^{10}_1,q^{10}_2,q^{10}_3)=(2,1,0)$ would represent an equally good
choice, provided we appropriately rescale the expansion parameter $\lambda$.
Turning to the $\bar 5$ charges, if we take \cite{lopsu1,lops2,lopsu5af,u1again,chargesnu}
\beq
(q^{\bar 5}_1,q^{\bar 5}_2,q^{\bar 5}_3)=(b,0,0)~~~~~~~~~~~~b\ge0~~~, 
\label{cha5}
\eeq
together with eq. (\ref{cha10}) we get the patterns $m_d:m_s:m_b\sim m_e:m_\mu:m_\tau \sim 
\lambda^{3+b}:\lambda^2:1$. Moreover,
the 22, 23, 32, 33 entries of the effective light neutrino mass matrix $m_\nu$ are all O(1), thus accommodating 
the nearly maximal value of $s_{23}$. The small non diagonal terms of the
charged lepton mass matrix cannot change this. 
We obtain, where arbitrary o(1) coefficients are omitted:   
\beq 
m_\nu=
\left(
\matrix{
\lambda^{2b}&\lambda^b&\lambda^b\cr
\lambda^b&1&1\cr
\lambda^b&1&1}\right){v_u^2\over \Lambda}~~~~~~~({\tt A,SA})~~~, 
\label{mlep}
\eeq 
where $v_u$ is the VEVs of the Higgs doublet giving mass to the up quarks and all the entries are specified 
up to order one coefficients. If we take $v_u\sim 250~{\rm GeV}$, the mass scale ${\Lambda}$ of
the heavy Majorana neutrinos turns out to be close to the unification scale, 
${\Lambda}\sim 10^{15}~{\rm GeV}$.

If $b$ vanishes, then the light neutrino mass matrix will be structure-less and we recover the anarchical (A)
picture of neutrinos \cite{anarchy}. In a large sample of anarchical models, generated with 
random coefficients, the resulting neutrino mass spectrum can exhibit either normal or inverse hierarchy. 
For down quarks and charged leptons we obtain a weakened hierarchy, essentially the square root than that of up quarks.

If $b$ is positive, then the light neutrino mass matrix will be structure-less only in the (2,3) sub-sector and we
get the so-called semi-anarchical (SA) models, defined by a matrix like in eq.(\ref{mlep}) with a 23 subdeterminant generically of order 1). 
In this case, the neutrino mass spectrum has normal hierarchy. However,
unless the (2,3) sub-determinant is accidentally suppressed, atmospheric and solar oscillation frequencies are 
expected to be of the same order and, in addition, the preferred solar mixing angle is small. Nevertheless, such a
suppression can occur in a fraction of semi-anarchical models generated with random, order one coefficients. The
real advantage over the fully anarchical scheme is represented by the suppression in $U_{e3}$. 

Note that in all previous cases we could add a constant to $q^{\bar 5}_i$, for example by taking 
$(q^{\bar5}_1,q^{\bar5}_2,q^{\bar5}_3)
=  (2+b,2,2)$. This would only have the consequence to leave the top quark as the only unsuppressed mass and to
decrease the resulting value of $\tan{\beta}=v_u/v_d$ down to $\lambda^2 m_t/m_b$. A constant shift of the charges $q^1_i$ might also provide a suppression
of the leading $\nu^c$ mass eigenvalue, from $\Lambda_c$ down to the 
appropriate scale $\Lambda$. One
can also consider models where the 5 and $\bar 5$ Higgs charges are different, as in the ``realistic'' SU(5) model of
ref. \cite{afm2}. Also in these models the top mass could be the only one to be non-vanishing in the symmetric limit and the
value of $\tan{\beta}$ can be adjusted.

\subsubsection{F(fermions) and F(flavons) of both signs}

Models with naturally large 23 splittings are obtained if we allow negative charges and, at the same time, either
introduce flavons of opposite charges or stipulate that matrix elements with overall negative charge are put to
zero. For example, we can assign to the fermion fields the set of
F charges given by:
\bea
(q^{10}_1,q^{10}_2,q^{10}_3) &= & (3,2,0) \nn\\
(q^{\bar 5}_1,q^{\bar 5}_2,q^{\bar 5}_3) &= & (b,0,0)~~~~~~~~~~~~b\ge 2a>0\nn\\
(q^{1}_1,q^{1}_2,q^{1}_3) &= & (a,-a,0) ~~~.\label{cha1}
\eea   
We consider the Yukawa coupling allowed by U(1)$_{\rm F}$-neutral  Higgs multiplets
in the $5$ and ${\bar 5}$ SU(5) representations and by a pair $\theta$ and
${\bar\theta}$ of SU(5) singlets with F$=1$ and F$=-1$, respectively. 
If $b=2$ or 3, the up, down and charged lepton sectors are
not essentially different than in the SA case. Also in this case the O(1) off-diagonal entry of $m_l$, typical
of lopsided models, gives rise to a large LH  mixing in the 23 block which corresponds to a large
RH mixing in the
$d$ mass matrix. In the neutrino sector, 
after diagonalization of the charged lepton sector and after
integrating out the heavy RH neutrinos we obtain the following neutrino mass matrix in the low-energy
effective theory:
\beq
m_\nu=
\left(
\matrix{
\lambda^{2 b}&\lambda^b&\lambda^b\cr
\lambda^b&1+\lambda^a{\lambda'}^a&1+\lambda^a{\lambda'}^a\cr
\lambda^b&1+\lambda^a{\lambda'}^a&1+\lambda^a{\lambda'}^a}\right)
{v_u^2\over {\Lambda}}~~~~~~~({\tt H}),
\label{mnu}
\eeq
where $\lambda'$ is given by $\langle\theta\rangle/\Lambda_c$ and ${\Lambda}$ as before denotes the large mass 
scale associated to the
RH neutrinos: ${\Lambda}\gg v_{u,d}$. 
The O(1) elements in the 23 block are produced by combining the
large  LH mixing induced by the charged lepton sector and the large LH mixing in $m_D$. A crucial
property of $m_\nu$ is that, as a result of the see-saw mechanism and of the specific U(1)$_{\rm F}$ 
charge assignment, the determinant of the 23 block is automatically of $O(\lambda^a{\lambda'}^a)$ 
(for this the presence of negative
charge values, leading to the presence of both $\lambda$ and
$\lambda'$ is essential \cite{lops2,lopsu5af}).  If we take
$\lambda\approx\lambda'$, it is easy to verify that the eigenvalues of $m_\nu$ satisfy  the relations:
\beq m_1:m_2:m_3  = \lambda^{2(b-a)}:\lambda^{2a}:1~~.
\eeq 
The atmospheric neutrino oscillations require 
$m_3^2\sim 10^{-3}~{\rm eV}^2$. The squared mass difference between the lightest states is  of
$O(\lambda^{4a})~m_3^2$, not far from the LA solution to the solar neutrino problem if we choose $a=1$. In general $U_{e3}$ is
non-vanishing, of $O(\lambda^b)$. Finally, beyond the large mixing in the 23 sector,
$m_\nu$  provides a mixing angle $\theta_{12} \sim \lambda^{b-2a}$ 
in the 12 sector. 
When $b=2 a$, as for instance in the case 
$b=2$ and $a=1$, the  LA solution can be reproduced and 
the resulting neutrino spectrum is hierarchical (H).

Alternatively, an inversely hierarchical (IH) spectrum can be obtained by choosing:
\bea
(q^{10}_1,q^{10}_2,q^{10}_3) &= & (3,2,0) \nn\\
(q^{\bar 5}_1,q^{\bar 5}_2,q^{\bar 5}_3) &= & (1,-1,-1)~~~~~~~~~~~~\nn\\
(q^{1}_1,q^{1}_2,q^{1}_3) &= & (-1,1,0) \nn\\
(q_{H_u},q_{H_d})&=&(0,1)~~~.\label{cha2}
\eea   
Due to the non-vanishing charge of the $H_d$ Higgs doublet, in the charged lepton sector
we recover the same pattern previously discussed.
The light neutrino mass matrix is given by:
\beq
m_\nu=
\left(
\begin{array}{ccc}
\lambda^2 & 1 & 1\\
1 & \lambda'^2 & \lambda'^2\\
1 & \lambda'^2 & \lambda'^2
\end{array}
\right)~~~~~~~~~~({\tt IH})~~~.
\label{mnuih}
\eeq
The ratio between the solar and atmospheric
oscillation frequencies is not directly related to the sub-determinant 
of the block 23, in this case.

\vspace{0.1cm}
\begin{table}[!h]
\caption{Models and their flavour charges. \label{updt_tab1}}
\vspace{0.4cm}
\begin{center}
\begin{tabular}{|c|c|c|c|c|}
\hline 
& & & & \\ 
{\tt Model}& ${{\Psi_{10}}}$ & ${\Psi_{\bar 5}}$ & ${{\Psi_1}}$ & ${(H_u,H_d)}$ \\ 
& & & & \\
\hline
\hline
 & & & & \\ 
{\tt Anarchical ($A$)}& (3,2,0)& (0,0,0) & (0,0,0) & (0,0)\\ 
& & & & \\
\hline
 & & & & \\ 
{\tt Semi-Anarchical ($SA$)}& (2,1,0) & (1,0,0) & (2,1,0) & (0,0) \\ 
& & & & \\
\hline
\hline 
& & & & \\
{\tt Hierarchical ($H_{I}$)}& (6,4,0)& (2,0,0) & (1,-1,0) & (0,0)\\ 
& & & & \\
\hline
& & & & \\
{\tt Hierarchical ($H_{II}$)}& (5,3,0)& (2,0,0) & (1,-1,0) & (0,0)\\ 
& & & & \\
\hline
& & & & \\ 
{\tt Inversely Hierarchical ($IH_{I}$)}& (3,2,0) & (1,-1,-1)& (-1,+1,0)& (0,+1) \\
& & & & \\
\hline& & & & \\ 
{\tt Inversely Hierarchical ($IH_{II}$)}& (6,4,0) & (1,-1,-1)& (-1,+1,0)& (0,+1) \\
& & & & \\
\hline
\end{tabular}
\end{center}
\end{table}

A representative set of models is listed in table 2. Note that in some cases the charges for $\Psi_{10}$ have been changed from $(3,2,0)$  (our reference values in eqs. (\ref{cha10}), (\ref{cha1}), and (\ref{cha2})) to $(6,4,0)$ or $(5,3,0)$. These values are a posteriori better suited the reproduce the moderate level of hierarchy implied by the present neutrino oscillation data. Since the neutrino mixing parameters are completely independent on
the 10 charges, this change is only important for a
better fit to quark and charged lepton masses and mixings once a rather large value of $\lambda$ is derived from the neutrino data.  
The hierarchical and the inversely hierarchical models may come into  several varieties depending on the number and
the charge of the flavour symmetry breaking (FSB) parameters. Above we have considered the case
of two (II)  oppositely charged flavons with symmetry breaking
parameters $\lambda$ and $\lambda'$. It may be noticed that the presence of two multiplets $\theta$ and
${\bar \theta}$ with opposite F charges could hardly be reconciled, without adding extra structure to the model,
with a large common VEV for these fields, due to possible analytic terms of the kind $(\theta {\bar \theta})^n$ in
the superpotential. Therefore it is instructive to explore the consequences of allowing only the negatively
charged ${\bar \theta}$ field in the theory, case I.
In case I, it is impossible to compensate negative F charges in the Yukawa
couplings and the corresponding entries in  the neutrino mass matrices vanish. Eventually these zeroes are filled by
small contributions, arising, for instance, from the diagonalization of the charged lepton sector or from the
transformations needed to make the kinetic terms canonical. 

Another important ingredient is represented by the see-saw mechanism \cite{seesaw}. Hierarchical models and
semi-anarchical  models have similar charges in the $(10,{\bar 5})$ sectors and, in the absence of the see-saw
mechanism, they would give rise to similar results. Even when the results are expected to be independent from the
charges of the RH neutrinos, as it is the case for the anarchical and semi-anarchical models, the see-saw
mechanism can induce some sizeable effect in a statistical  analysis. For this reason, for each type of model, but
the normal-hierarchical ones (the mechanism for the 23 sub-determinant suppression is in fact based on the see-saw
mechanism), it is interesting to study the case where RH neutrinos are present and the see-saw contribution 
is the dominant one (SS) and the case where they are absent and the mass matrix is saturated by the 
non-renormalizable contribution (NOSS). 

With this classification in mind, we can distinguish the following type of models, all
supported by specific choices of U(1) charges:
${\rm A_{SS}}$, ${\rm A_{NOSS}}$, ${\rm SA_{SS}}$, ${\rm SA_{NOSS}}$,
${\rm H_{(SS,I)}}$, ${\rm H_{(SS,II)}}$, ${\rm IH_{(SS,I)}}$, 
${\rm IH_{(SS,II)}}$, ${\rm IH_{(NOSS,I)}}$ and ${\rm IH_{(NOSS,II)}}$.

\begin{figure}[!h]
\centerline{   \psfig{file=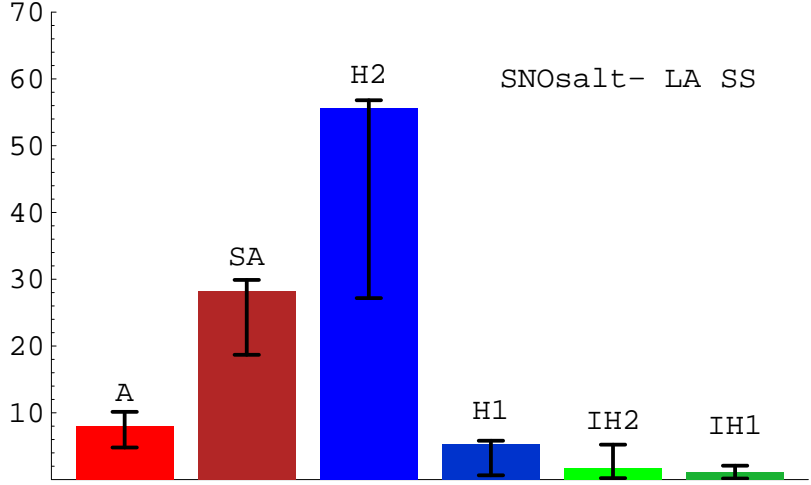,width=0.7 \textwidth}  }
\caption{Relative success rates for the LA solution, with see-saw. 
The sum of the rates has been normalized to 100. The results correspond to the 
default choice ${\cal I}=[0.5,2]$, and to the following values 
of $\lambda=\lambda'$: $0.2$, $0.25$, $0.35$, $0.45$, $0.45$, $0.25$ 
for the models ${\rm A_{SS}}$, ${\rm SA_{SS}}$, ${\rm H_{(SS,II)}}$, 
${\rm H_{(SS,I)}}$, ${\rm IH_{(SS,II)}}$ and ${\rm IH_{(SS,I)}}$, 
respectively. The error bars represent
the linear sum of the systematic error due to the
choice of ${\cal I}$ and the statistical error (see text).}
\label{updt_barlass}
\end{figure}

\begin{figure}[!h]
\vskip .5 cm
\centerline{  \psfig{file=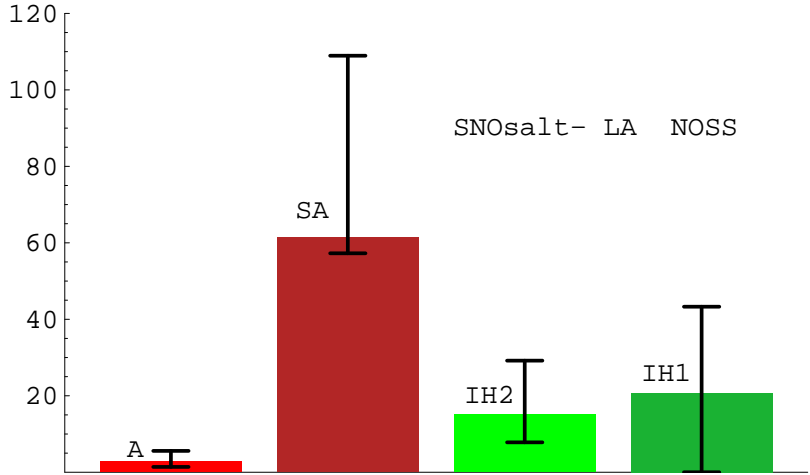,width=0.7 \textwidth}   }
\caption{Relative success rates for the LA solution, without see-saw. 
The sum of the rates has been normalized to 100. The results correspond to the 
default choice ${\cal I}=[0.5,2]$, and to the following values 
of $\lambda=\lambda'$: $0.2$, $0.2$, $0.25$, $0.25$
for the models ${\rm A_{NOSS}}$, ${\rm SA_{NOSS}}$, ${\rm IH_{(NOSS,II)}}$, 
and ${\rm IH_{(NOSS,I)}}$, respectively (in our notation there are no ${\rm H_{(NOSS,I)}}$, 
${\rm H_{(NOSS,II)}}$ models). 
The error bars represent the linear sum of the systematic error due to the
choice of ${\cal I}$ and the statistical error (see text).}
\label{updt_barlanoss}
\end{figure}

It is interesting to quantify the ability of each model in reproducing the observed oscillation
parameters. For anarchy, it has been observed that random generated, order-one 
entries of the neutrino mass matrices (in appropriate units), correctly fit the experimental data
with a success rate of few percent. It is natural to extend this analysis to include also
the other models based on SU(5) $\times$ U(1) \cite{afm03}, which have mass matrix elements defined up to 
order-one dimensionless coefficients $y_{ij}$ (see eq. \ref{m1}). For each model, successful 
points in parameter space are selected by asking that the four observable quantities $O_1=r
\equiv \Delta m^2_{12}/\vert\Delta m^2_{23}\vert$,
$O_2=\tan^2\theta_{12}$, $O_3=\vert U_{e3}\vert\equiv\vert\sin\theta_{13}\vert$ and 
$O_4=\tan^2\theta_{23}$ fall in the approximately 3$\sigma$ allowed ranges \cite{Valle}: 
\beq
\begin{array}{l}
 0.018 < r < 0.053\\
\vert U_{e3} \vert < 0.23\\ 0.30 <\tan^2\theta_{12}<0.64\\ 0.45 <\tan^2\theta_{23}<2.57
\end{array}
\label{lanew}
\eeq
The coefficients $y_{ij}$ of the neutrino sector are random complex numbers with absolute values
and phases uniformly distributed in intervals ${\cal I}=[0.5,2]$ and $[0,2\pi]$ respectively.
The dependence of the results on these choices can be estimated by varying
${\cal I}$.
For each model an optimization procedure selects the value of the flavour symmetry breaking
parameter $\lambda=\lambda'$ that maximizes the success rate.
The success rates are displayed in figs. \ref{updt_barlass} and  \ref{updt_barlanoss}, separately for the SS and NOSS
cases. The two sets of models have been individually normalized to give a total rate 100. 
From the 
histograms in figs. \ref{updt_barlass} and  \ref{updt_barlanoss}  we see that normal hierarchy models (with two oppositely charged flavons ${\rm H_{II}}$) are neatly
preferred over anarchy and inverse hierarchy in the context of these SU(5)$\times$U(1) models. In particular, in the SS
case, the ${\rm H_{II}}$ models with normal hierarchy and suppressed 23 sub-determinant are clearly preferred. Models of the type ${\rm H_{I}}$ are disfavoured because they tend to give $\tan{\theta_{12}}^2>1$.
We recall that for the chosen charge values the ${\rm H_{II}}$ model is of the lopsided type. In the NOSS case the see-saw suppression of the 23 determinant is clearly not operative and all normal hierarchy models coincide with SA.

\vskip 0.3cm

An interesting question is whether the disfavouring of IH models that we find in our SU(5)$\times$U(1) framework can be
extended to a more general context. In the limit of vanishing $\lambda$ and $\lambda'$ the IH texture (see eq. (\ref{mnuih})) 
becomes close to that of
bimaximal mixing and $\theta_{13}=0$ (actually with $r=0$). In our U(1) models $r\approx
\vert U_{e3}\vert \approx\vert\tan^2\theta_{12}-1\vert \approx O(\lambda^2)$ (for $\lambda=\lambda'$). In particular the charged lepton mixings
cannot displace too much $\theta_{12}$ from its maximal value because the small value of the electron mass forces a
sufficiently large value of the relevant charges, which in turn implies that the charged lepton mixing correction to
$\theta_{12}$ is small. We have already mentioned that corrections from the charged lepton sector can in principle bring the predictions of a neutrino matrix of the bimixing type in agreement with the data and that the smallness of $s_{13}$ induces strong constraints. In the particular setup of $U(1)_F$ models we have seen that charged lepton corrections are too small to make the solar angle sufficiently different from maximal.

In conclusion, models based on SU(5) $\times$ U(1)$_{\rm F}$ are clearly toy models that can only aim at a semiquantitative
description of fermion masses. In fact only the order of magnitude of each matrix entry can be specified. However
it is rather impressive that a reasonable description of fermion masses, now also including neutrino masses and
mixings, can be obtained in this simple context, which is suggestive of a deeper relation between gauge and flavour
quantum numbers. Moreover, all possible type of mass hierarchies can be reproduced within this framework.
In a statistically based comparison, the range of $r$ and the small upper
limit on
$U_{e3}$  are sufficiently constraining to make anarchy neatly disfavoured with respect to models
with built-in hierarchy. If only neutrinos are considered, one might counterargue that hierarchical models have
at least one more parameter than anarchy, in our case the parameter $\lambda$. However, if one looks at quarks and
leptons together, as in the GUT models that we consider, then the same parameter that plays the role of an order parameter
for the CKM matrix, for example, the Cabibbo angle, can be successfully used to reproduce also the hierarchy
implied by the present neutrino data. Actually it is interesting that the data now favour a moderate hierarchy, well described in terms of the moderately small Cabibbo angle.

\section{Neutrino Mixings from the Charged Lepton \\Sector}

The observed neutrino mixing matrix $U=U_e^\dagger U_{\nu}$, in the limit of vanishing $\sin{\theta_{13}}=s_{13}$, has the
approximate form:
\be U=
\left[\matrix{
c&s&0\cr s/\sqrt{2}&-c/\sqrt{2}&1/\sqrt{2}\cr -s/\sqrt{2}&c/\sqrt{2}&1/\sqrt{2}} 
\right]~~~~~, 
\label{Uobs}
\ee
where $s$ and $c$ stand for $\sin{\theta_{12}}$ and $\cos{\theta_{12}}$ respectively and we took the atmospheric angle
$\theta_{23}$ as exactly maximal. The effective mass matrix of light neutrinos is in general given by:
\be
m_\nu= U^* m_{\nu}^{diag}U^\dagger~.~~~~~~\label{mnu}\\
\ee
Starting from the lagrangian basis, where all symmetries of the theory are specified, we want to investigate whether it is possible to obtain the observed mixings in a natural way from the diagonalization of the charged
lepton mass matrix by $U_e$ while $U_{\nu}$ is nearly diagonal. The possible deviations from maximal $\theta_{23}$ and from $s_{13}=0$ can be
omitted in eq. (\ref{Uobs}) and attributed to small effects from $U_{\nu}$ that will be in general not exactly zero. One
might think that given the rather symmetric role of $U_e$ and $U_{\nu}$ in the formula $U=U_e^\dagger U_{\nu}$ one way or the
other should be equivalent. But we will show that this is not so. Actually now that we know that also the solar angle
$\theta_{12}$ is large, this tends to clash with a small $\theta_{13}$, in the case of mixings dominated by $U_e$. 

In terms of $U_e$ the charged lepton mass matrix $m_e$ (defined as $\bar R m_eL$ from right-handed ($R$) and left-handed
($L$) charged lepton fields in the lagrangian basis) can be written as:
\be
m_e=V_e m_e^{diag} U_e^\dagger~~.~~~~~~\label{me}\\
\ee
Indeed $L_{diag}= U_e L$ and $R_{diag}=V_e R$ are the transformations between the lagrangian and the mass basis for the $R$ and $L$
fields. Assuming that $U\sim U_e^\dagger$, given that $m_e^{diag}=Diag[m_e,m_\mu,m_\tau]$ we can write:
\be
m_e=V_e m_e^{diag} U= V_e \left[\matrix{
cm_e&sm_e&0\cr s/\sqrt{2}m_\mu&-c/\sqrt{2}m_\mu&m_\mu/\sqrt{2}\cr -s/\sqrt{2}m_\tau&c/\sqrt{2}m_\tau&m_\tau/\sqrt{2}} 
\right]~~~~~.\label{meV}
\ee
We will come back later on the matrix $V_e$ that determines the right-handed mixings of charged leptons. For the time being
it is already interesting to consider the matrix $m_e^\dagger m_e$ which is completely fixed by $U_e$: 
\be
m_e^\dagger m_e=U_e(m_e^{diag})^2 U_e^\dagger~~.
\label{me2}\\
\ee
Neglecting for simplicity the electron mass, we find, for $U_e^\dagger=U$:
\be
m_e^\dagger m_e= U^\dagger(m_e^{diag})^2 U = \frac{1}{2}(m_\tau^2+m_\mu^2)\left[\matrix{
s^2&-cs&-s(1-2\lambda^4)\cr -cs&c^2&c(1-2\lambda^4)\cr
-s(1-2\lambda^4)&c(1-2\lambda^4)&1} 
\right]~~~~~,\label{me22}
\ee
where we defined
\be
\frac{m_\tau^2-m_\mu^2}{m_\tau^2+m_\mu^2} = 1-2\lambda^4~~~~~\label{eta}\\
\ee
so that approximately $\lambda^4 \sim m_{\mu}^2/m_{\tau}^2$. The problem with this expression for $m_e^\dagger m_e$ is that all
matrix elements are of the same order and the vanishing of $s_{13}$ as well as the hierarchy of the eigenvalues arise from
precise relations among the different matrix elements. For example, the result $s_{13}=0$ is obtained because the
eigenvector with zero eigenvalue is of the form $e_1=(c,s,0)^T$ and the crucial zero is present because 
the first two columns are proportional in eq. (\ref{me22}). These features are more difficult to implement in a
natural way than matrices with texture zeros or with a hierarchy of matrix elements. Only if the solar angle $\theta_{12}$
is small, that is $s$ is small, then the first row and column are nearly vanishing and $s_{13}$ is automatically small.

Consider, for comparison, the case where we do not make the
hypothesis that all the mixings are generated by the charged leptons, but rather that $U_e \sim 1$. 
To make the comparison more direct, let us assume
that the neutrino mass spectrum is of the normal hierarchy type with dominance of
$m_3$: $m_\nu^{diag}\sim m_3 Diag[0,\xi^2,1]$, where $\xi^2=m_2/m_3$ is small and $m_1$ is neglected. In this case, the
effective light neutrino mass matrix is given by (note the crucial transposition of $U$, 
which in eq. (\ref{Uobs}) is real,  with respect to eq. (\ref{me22})):
\be
m_\nu = U^*  m_\nu^{diag} U^\dagger \sim \frac{m_3}{2}\left[\matrix{
s^2\xi^2&-cs\xi^2/\sqrt{2}&cs\xi^2/\sqrt{2}\cr -cs\xi^2/\sqrt{2}&(1+c^2\xi^2)/2&(1-c^2\xi^2)/2\cr
cs\xi^2/\sqrt{2}&(1-c^2\xi^2)/2&(1+c^2\xi^2)/2} 
\right]~~~~~.\label{menu}
\ee
In this case, no matter what the value of $s$ is, the first row and column are of order $\xi^2$. By replacing terms of
order $\xi^2$ by generic small terms of the same order, $s_{13}$ remains of order $\xi^2$. We can also replace the
terms of order 1 in the 23 sector by generic order 1 quantities provided that we have a natural way of guaranteeing that the
subdeterminant 23 is suppressed and remains of order $\xi^2$. As well known this suppression can be naturally induced
through the see-saw mechanism either by dominance of a single right-handed Majorana neutrino \cite{dominance} or by a lopsided \cite{lops}, \cite{lopsided} neutrino Dirac
matrix. Natural realizations of this strategy have been constructed, for example, in the context of U(1)$_F$ flavour models \cite{lopsu5af}, \cite{afm2}, \cite{lopsu1}.

We now come back to the expression for the charged lepton mass matrix $m_e$ in eq. (\ref{meV}) where the matrix $V_e$ appears.
This matrix describing the right-handed mixings of charged leptons is not related to neutrino mixings. In minimal SU(5) the
relation $m_e=m_d^T$ holds between the charged lepton and the down quark mass matrices. In this case $V_e$ describes the
left-handed down quark mixings: $V_e=U_d$. The CKM matrix, as well known, is given by $V_{CKM}=U_u^\dagger U_d$. Given that
the quark mixing angles are small, either both $U_u$ and $U_d$ are nearly diagonal or they are nearly equal. Thus one
possibility is that $U_d$ is nearly diagonal. In this case, for $V_e=U_d$, $m_e$ is approximately given by eq. (\ref{meV}) with $V_e \sim 1$.
Neglecting the electron mass and setting $\lambda^2=m_{\mu}/m_{\tau}$ we obtain:
\be
m_e\approx m_e^{diag} U= m_{\tau}\left[\matrix{
0&0&0\cr s/\sqrt{2}\lambda^2&-c/\sqrt{2}\lambda^2&\lambda^2/\sqrt{2}\cr -s/\sqrt{2}&c/\sqrt{2}&1/\sqrt{2}} 
\right]~~~~~.\label{meV1}
\ee
This matrix is a generalization of lopsided models with all three matrix elements in the third row of order 1 (unless $s$ is
small: for small solar angle we go back to the situation of normal lopsided models). We recall that lopsided models
with the 23 and 33 matrix elements of order 1 provide a natural way to understand a large 23 mixing angle. In fact from the
matrix relation
\be
\left[\matrix{
0&0&0\cr 0&0&0\cr 0&s_{23}&c_{23}\cr} 
\right]\left[\matrix{
1&0&0\cr 0&c_{23}&s_{23}\cr 0&-s_{23}&c_{23}\cr} 
\right]= \left[\matrix{
0&0&0\cr 0&0&0\cr 0&0&1\cr} 
\right]~~~~~,\label{lop}
\ee
we see that in lopsided models one automatically gets a large 23 mixing from $U_e$. In the generalized case of
eq. (\ref{meV1}), while the natural prediction of a large 23 mixings remains, the relation $s_{13}=0$ does not arise
automatically if the entries of the matrix are replaced by generic order 1 terms in the third row and of order $\lambda^2$
in the second row. If we call $v_3$ the 3-vector with components of order 1 in the third row and $\lambda^2 v_\lambda$ the
vector of the second row, we can easily check that to obtain $s_{13}=0$ 
it is needed that both $v_\lambda$ and $v_3$ are orthogonal to a vector of 
the form $(c,s,0)$.

In democratic models all matrices $U_u$, $U_d$, $U_e$ are nearly equal and the smallness of quark mixings
arises from a compensation between $U_u^\dagger$ and $U_d$. This sort of models correspond, for $V_e=U_d$, to $V_e=U_e=U^\dagger$ and a symmetric
matrix $m_e$: $m_e=U^\dagger m_e^{diag} U$. In this case we obtain a matrix exactly equal to that in eq. (\ref{me22}) for $m_e^\dagger
m_e$ except that squared masses are replaced by masses. As discussed in the case of eq. (\ref{me22}), 
we need fine-tuning in order to reproduce the observed hierarchy of mass and to obtain
$s_{13}=0$ unless the solar angle $s$ is small. Note in fact, that in the democratic model of \cite{fldem},
the vanishing of $s_{13}$ is only accommodated but not predicted.

\subsection{A Natural Class of Models}

We now attempt to identify a set of conditions that make possible the
construction of an explicit model where the mixing in the lepton sector
is dominated by the charged lepton contribution.
One obvious condition is a dynamical or a symmetry principle that forces
the light neutrino mass matrix to be diagonal in the lagrangian basis.
The simplest flavour symmetries cannot fulfill this requirement in a 
simple way.
For instance, a U(1) symmetry can lead to a nearly diagonal neutrino mass
matrix, of the form:
\be 
m_\nu=
\left[\matrix{
\xi^{2p}&\xi^{p+1}&\xi^p\cr 
\xi^{p+1}&\xi^2&\xi\cr 
\xi^p&\xi&1} 
\right] m~~~~~, 
\label{mnund}
\ee
where $\xi<1$ is a U(1) breaking parameter, $p\ge 1$
and all matrix elements are defined up to unknown order one coefficients.
The problem with this matrix is that the ratio between the solar and the 
atmospheric squared mass differences, close to 1/40, is approximately
given by $\xi^4$ and, consequently,
a large atmospheric mixing angle is already induced by $m_\nu$ itself.
If we consider a discrete symmetry like $S_3$, $m_\nu$ can
be of the general form:
\be 
m_\nu=
\left[\matrix{
1&0&0\cr 
0&1&0\cr 
0&0&1} 
\right] m +
\alpha \left[\matrix{
0&1&1\cr 
1&0&1\cr 
1&1&0} 
\right] m
~~~, 
\label{mnus3}
\ee
where $\alpha$ is an arbitrary parameter. In this case we need 
both the extra assumption $|\alpha|\ll 1$ and a specific symmetry breaking 
sector to lift the mass degeneracy \cite{fldem}.
A stronger symmetry like O(3) removes from eq. (\ref{mnus3}) the non-diagonal
invariant, but requires a non-trivial symmetry breaking sector
with a vacuum alignment problem in order to keep the
neutrino sector diagonal while allowing large off-diagonal terms
for charged leptons \cite{barbierio3}. 

A simple, though not economical, possibility to achieve a diagonal
neutrino mass matrix, is to introduce three independent U(1) symmetries,
one for each flavour \cite{afm04}: 
F=U(1)$_{F_1}\times$U(1)$_{F_2}\times$U(1)$_{F_3}\times$..., where F denotes the flavour symmetry group.
The Higgs doublet giving mass to the up-type quarks is neutral under F.
Each lepton doublet is charged under a different U(1) factor, with the same
charge +1. In the symmetric phase all neutrinos are exactly massless.
Flavour symmetry breaking is obtained by non-vanishing vacuum expectation 
values (VEVs) of three 
flavon fields, also charged under a separate U(1) factor, with charge -2. 
In this way only diagonal neutrino mass terms are induced. 
If the VEVs in the flavon sector
are similar, we expect neutrino masses of the same order, and 
the observed hierarchy between the solar and atmospheric squared mass
differences requires a modest adjustment of the flavon vev's and/or of the coefficients of 
the lepton violating operators.

A second condition can be identified by considering a mass matrix 
for the charged leptons which is very close, but slightly more 
general than the one of eq. (\ref{meV1}):
\be
m_e=\left[\matrix{
O(\lambda^4)&O(\lambda^4)&O(\lambda^4)\cr 
x_{21}\lambda^2& x_{22}\lambda^2&O(\lambda^2)\cr 
x_{31}&x_{32} &O(1)} 
\right] m~~~~~,
\label{meV2}
\ee
where $x_{ij}$ $(i=2,3)$ $(j=1,2)$ is a matrix of order one coefficients
with vanishing determinant:
\be
x_{21} x_{32} - x_{22} x_{31}=0~~~~~.
\label{det0}
\ee
The eigenvalues of $m_e$ in units of $m$ are of order 1, 
$\lambda^2$ and $\lambda^4$, as required by the charged lepton masses.
Moreover, the eigenvalue of order $\lambda^8$ of $m_e^\dagger m_e$ 
has an eigenvector:
\be
(c,s,O(\lambda^4))~~~~~~\frac{s}{c}=-\frac{x_{31}}{x_{32}}+O(\lambda^4)~~~~~.
\ee
In terms of the lepton mixing matrix $U=U_e^\dagger$, this means 
$\theta_{13}=O(\lambda^4)$ and $\theta_{12}$ large, if 
$x_{31}\approx x_{32}$. When the remaining, unspecified parameters in $m_e$
are all of order one, also the atmospheric mixing angle $\theta_{23}$
is large. 
Notice that, by neglecting $O(\lambda^4)$ terms, the following relation
holds for the mass matrix $m_e/m$ (cfr. eq. (\ref{lop})):
\be
\left[\matrix{
0&0&0\cr 
x_{21}\lambda^2& x_{22}\lambda^2&O(\lambda^2)\cr 
x_{31}&x_{32} &O(1)} 
\right]
\left[\matrix{
c&-s&0\cr 
s&c&0\cr 
0&0&1} 
\right]=
\left[\matrix{
0&0&0\cr 
0& \sqrt{x_{21}^2+x_{22}^2}\lambda^2&O(\lambda^2)\cr
0&\sqrt{x_{31}^2+x_{32}^2} &O(1)} 
\right]~~~~~,
\label{meV3}
\ee
where
\be
\frac{s}{c}=-\frac{x_{31}}{x_{32}}~~~~~.
\ee
Therefore the natural parametrization of the unitary matrix
$U_e$ that diagonalizes $m_e^\dagger m_e$ in this approximation is:
\be
U_e={U^e}_{12} {U^e}_{23}~~~~~,
\ee
where $U^e_{ij}$ refers to unitary transformation in the $ij$ plane.
Using $U=U_e^\dagger$, we automatically find the leptonic mixing matrix 
in the standard parametrization $U=U_{23} U_{13} U_{12}$ (neglecting phases),
with $U_{23}={U^e}_{23}^\dagger$, $U_{13}=1$ and $U_{12}={U^e}_{12}^\dagger$.
Had we used the standard parametrization also for $U_e$, we would have
found three non-vanishing rotation angles $\theta^e_{ij}$ with non-trivial
relations in order to reproduce $\theta_{13}=0$.

This successful pattern of $m_e$, eq. (\ref{meV2}), has two features. The first one
is the hierarchy between the rows. It is not difficult to obtain this
in a natural way. For instance, we can require a U(1) flavour symmetry
acting non-trivially only on the right-handed charged leptons, thus
producing the required suppressions of the first and second rows.
The second one is the vanishing determinant condition of eq. (\ref{det0}).
We can easily reproduce this condition by exploiting a see-saw
mechanism operating in the charged lepton sector.

To show this we add to the field content
of the standard model additional vector-like fermion pairs 
$(L_a,L^c_a)$ $(a=1,...n)$ of SU(2) doublets, with hypercharges 
$Y=(-1/2,+1/2)$. The Lagrangian in the charged lepton sector
reads:
\be
{\cal L}={kinetic~terms}
+ \eta_{ij} e^c_i l_j h_d 
+ \lambda_{ia} e^c_i L_a h_d 
+ \mu_{aj} L^c_a l_j 
+ M_a L^c_a L_a + h.c.
\label{lag}
\ee
where $l_i$ and $e^c_i$ $(i=1,2,3)$ are the standard model leptons,
doublet and singlet under SU(2), respectively, and $h_d$ denotes
the Higgs doublet. We assume a diagonal mass matrix for the extra
fields. We expect $M_a,\mu_{aj}\gg \langle h_d\rangle$
and in this regime there are heavy fermions that can be
integrated out to produce a low-energy effective Lagrangian.
The heavy combinations are $L^c_a$ and 
\be
L_a+\frac{\mu_{aj}}{M_a} l_j~~~~~(a=1,...n)~~~.
\label{heavy}
\ee
These fields are set to zero by the equations of motion
in the static limit and we should express all remaining fermions
in term of the three combinations that are orthogonal to those
in eq. (\ref{heavy}), and which we identify with the light degrees of
freedom. To illustrate our point it is sufficient to work in the regime
$\vert\mu_{aj}\vert<\vert M_a\vert$ and expand the relevant quantities
at first order in $\vert\mu_{aj}/M_a\vert$.
To this approximation
the light lepton doublets $l'_i$ are:
\be
l'_i=l_i-\frac{\mu_{ai}}{M_a} L_a~~~~~.
\ee
The effective lagrangian reads:
\be
{\cal L}={kinetic~terms}
+ (\eta_{ij}-\frac{\lambda_{ia}\mu_{aj}}{M_a})~ e^c_i l'_j h_d 
+ h.c.~~~~~,
\label{lagr}
\ee
and the mass matrix for the charged leptons is:
\be
m_e=\langle h_d\rangle(\eta_{ij}-
\frac{\lambda_{ia}\mu_{aj}}{M_a})~~~~~~.
\ee
This result is analogous to what obtained in the neutrino sector
from the see-saw mechanism. There is a term in $m_e$
coming from the exchange of the heavy fields $(L_a,L^c_a)$, which
play the role of the right-handed neutrinos, and there is another
term that comes from a single operator and that cannot be interpreted
as due to the exchange of heavy modes. In the regime 
$1>\vert\mu/M\vert>\vert\eta/\lambda\vert$ the ``see-saw'' contribution dominates.
Moreover, if the lower left block in $m_e$
is dominated by a single exchange, for instance by $(L_1,L^c_1)$,
then
\be
\left[\matrix{
{m_e}_{21}&{m_e}_{22}\cr 
{m_e}_{31}&{m_e}_{32}}
\right]=\frac{\langle h_d\rangle}{M_1}
\left[\matrix{
\lambda_{21}\mu_{11}&\lambda_{21}\mu_{12} \cr 
\lambda_{31}\mu_{11}&\lambda_{31}\mu_{12}}
\right]~~~~~,
\ee
and the condition of vanishing determinant in eq. (\ref{det0})
is automatically satisfied.

Additional vector-like leptons are required by several extensions
of the standard model. For instance, a grand unified theory based
on the E$_6$ gauge symmetry group with three generations
of matter fields described by three 27 representations of E$_6$,
includes, beyond the standard model fermions, two SU(5) singlets
and an SU(5) vector-like $(5,\bar{5})$ pair per each generation.
In such a model a ``see-saw'' mechanism induced by the exchange
of heavy $(5,\bar{5})$ fields is not an option, but a necessary
ingredient to recover the correct number of light degrees of freedom.
We should still show that it is possible to combine the above 
conditions in a natural and consistent framework. In ref.\cite{uslast}
we have presented, as an existence proof, a supersymmetric 
SU(5) grand unified model possessing a flavour symmetry 
F=U(1)$_{F_0}\times$U(1)$_{F_1}\times$U(1)$_{F_2}\times$U(1)$_{F_3}$.
The first U(1)$_{F_0}$ factor is responsible for the hierarchy of masses
and mixing angles in the up-type quark sector as well as for the
hierarchy between the rows in the charged lepton mass matrix. 
The remaining part of F guarantees a diagonal
neutrino mass matrix and, at the same time, dominance of a single
heavy $(5,\bar{5})$ pair in the lower left block of $m_e$.
Notice that, at variance with most of the other existing models
\cite{afrev}, this framework
predicts a small value for $\theta_{13}$, of order $\lambda^4$
which is at the border of sensitivity of future neutrino 
factories. 


\subsection{Corrections to Bimixing from $U_e$}

Even when the neutrino mass matrix $U_\nu$ is not diagonal in the lagrangian 
basis, the contribution
from the charged lepton sector can be relevant or even crucial to
reproduce the observed mixing pattern. An important example arises if 
the neutrino matrix $U_\nu$ instead of being taken as nearly diagonal, 
is instead assumed of a particularly simple form, like for bimixing: 
\be 
U_\nu=
\left[\matrix{
1/\sqrt{2}&1/\sqrt{2}&0\cr 1/2&-1/2&1/\sqrt{2}\cr -1/2&1/2&1/\sqrt{2}} 
\right]~~~~~. 
\label{Ubim}
\ee
This configuration can be obtained, for instance,
in inverse hierarchy models with a $L_e-L_\mu-L_\tau$ U(1) symmetry, 
which predicts maximal $\theta_{12}^\nu$, large $\theta_{23}^\nu$, vanishing $\theta_{13}^\nu$
and $\Delta m^2_{sol}=0$. 
After the breaking of this symmetry, the degeneracy between the first two neutrino 
generations is lifted and the small observed value of $\Delta m^2_{sol}$ can be easily reproduced.
Due to the small symmetry breaking parameters,
the mixing angles in eq. (\ref{Ubim}) also receive corrections, 
whose magnitude turns out \cite{lms2} to be controlled by $\Delta m^2_{sol}/\Delta m^2_{atm}$: 
$\theta_{13}^\nu \lappeq 1 - \tan^2 \theta_{12}^\nu \sim \Delta m^2_{sol}/(2\Delta m^2_{atm}) \sim 0.01$. 
These corrections are too small to account for the measured value of the solar angle. 
Thus, an important contribution from $U_e$ is necessary to reconcile bimixing with observation.

In this section we will reconsider the question of whether the observed 
pattern can result from the corrections induced by the charged lepton sector. 
Though not automatic, this appears to be at present a rather natural possibility \cite{chlept} 
- see also the recent detailed analysis of Ref. \cite{Frampton}.
Our aim is to investigate the impact of planned experimental improvements, 
in particular those on $|U_{e3}|$, on bimixing models.
To this purpose it is useful to adopt a convenient parametrization 
of mixing angles and phases.
Let us define
\be
\tilde U~=~ 
\left(\matrix{1&0&0 \cr 0&c_{23}&s_{23}\cr0&-s_{23}&c_{23}     } 
\right)
\left(\matrix{c_{13}&0&s_{13}e^{i\delta} \cr 0&1&0\cr -s_{13}e^{-i\delta}&0&c_{13}     } 
\right)
\left(\matrix{c_{12}&s_{12}&0 \cr -s_{12}&c_{12}&0\cr 0&0&1     } 
\right)~~~,
\label{ufi}
\ee 
where all the mixing angles belong to the first quadrant and $\delta$ to $[0,2 \pi]$.
The standard parameterization for $U$ reads: $U = \tilde U \times$ a diagonal $U(3)$ matrix 
accounting for the two Majorana phases of neutrinos (the overall phase is not physical).
Since in the following discussion we are not interested in the Majorana phases, 
we will focus our attention on $\tilde U$. 

It would be appealing to take the parameterization (\ref{ufi}) separately for $U_e$ and $U_\nu$, 
by writing $s_{12}$, $s^e_{12}$, $s^\nu_{12}$ etc to distinguish the mixings of the 
$U$, $U_e$ and $U_\nu$ matrices, respectively. 
However, as discussed in ref. \cite{afm04}, even disregarding Majorana phases,
$U$ is not just determined in terms of $\tilde U_e$ and $\tilde U_\nu$, with the latter 
defined to be of the form (\ref{ufi}). 
The reason is that, by means of field redefinitions $U_e$ and $U_\nu$ can be separately but 
{\it not simultaneously} written respectively 
as $\tilde U_e$ and $\tilde U_\nu \times$ a diagonal $U(3)$ matrix.
Without loss of generality we can adopt the following form for $U$:
\be
U=U_e^\dagger U_\nu = \underbrace{ \tilde U_e^\dagger  
{\rm diag}(-e^{- i (\alpha_1 + \alpha_2)}, - e^{-i \alpha_2},1)  \tilde U_\nu }_{= \bar U}
\times {\rm phases}
\label{Ugen}
\ee
where $\tilde U_e$, $\tilde U_\nu$ have the form (\ref{ufi}), the phases $\alpha_1$, $\alpha_2$ run 
from $0$ to $2 \pi$ and 
we have introduced two minus signs in the diagonal matrix for later convenience. 
This expression for $\bar U$ is not due to the Majorana nature of neutrinos and a similar result 
would also hold for quarks.

Assume now that $\tilde U_\nu$ corresponds to bimixing: $s^\nu_{13}=0$,
$s^\nu_{12}=c^\nu_{12}=1/\sqrt{2}$ and $s^{\nu}_{23}=c^{\nu}_{23}=1/\sqrt{2}$. 
Clearly, our discussion holds true irrespectively of the light neutrino spectrum. 
It is anyway instructive to explicitate the mass matrices, 
e.g. in the case of inverted hierarchy 
\be
m_\nu= 
\left(\matrix{0&1&1 \cr 1&0&0\cr1&0&0     } 
\right)\frac{\Delta m^2_{atm}}{\sqrt{2}}~,
\quad 
m_e= V_e  \left(\matrix{ 
m_e e^{-i(\alpha_1 + \alpha_2)}& - s^e_{12} m_e e^{-i \alpha_2}& - s^e_{13} m_e e^{i \delta_e}  \cr 
s^e_{12} m_\mu e^{-i(\alpha_1 + \alpha_2)}& m_\mu e^{-i \alpha_2} & - s^e_{23} m_\mu \cr 
s^e_{13} m_\tau e^{-i(\alpha_1 + \alpha_2 + \delta_e)}& s^e_{23} m_\tau e^{-i \alpha_2} &m_\tau   } 
\right)
\label{}
\ee 
where we have set $\Delta m^2_{sol} = 0$ since, as already mentioned, the 
corrections induced by setting it to the measured value are negligible in the
present discussion.

We then expand $\bar U$ of eq. (\ref{Ugen}) at first order in the small mixings of $\tilde U_e$,
$s^e_{12},$ $s^e_{13}$ and $s^e_{23}$
\footnote{To this approximation any ordering of the three 
small rotations in $U_e$ gives exactly the same results,
and our conclusions are independent on the adopted parametrization.}:
\bea
\bar U_{11}&=&- \frac{e^{-i(\alpha_1 + \alpha_2)}}{\sqrt{2}}
                - \frac{s^e_{12} e^{-i \alpha_2} + s^e_{13} e^{i \delta_e}}{2}\nn\\
\bar U_{12}&=&- \frac{e^{-i(\alpha_1 + \alpha_2)}}{\sqrt{2}}
                + \frac{s^e_{12} e^{-i \alpha_2} + s^e_{13} e^{i \delta_e}}{2}\nn\\
\bar U_{13}&=&  \frac{s^e_{12} e^{-i \alpha_2} - s^e_{13} e^{i \delta_e}}{\sqrt{2}}\nn\\
\bar U_{23}&=& - e^{-i \alpha_2}\frac{ 1 + s^e_{23} e^{i \alpha_2}  }{\sqrt{2}}\nn\\
\bar U_{33}&=&   \frac{1 - s^e_{23} e^{-i \alpha_2} }{\sqrt{2}}~~~.~~~~\label{UeUnBm}
\eea
The smallness of the observed $s_{13}$ implies that both $s^e_{12}$ and $s^e_{13}$ must be 
at most of order $s_{13}$. 
As a consequence, the amount of the deviation of $s_{12}$ from $1/\sqrt{2}$ is 
limited from the fact that it is generically of the same order as $s_{13}$.
Note that, instead, the deviation of the atmospheric angle $s_{23}$ from $1/\sqrt{2}$
is of second order in $s^e_{12}$ and $s^e_{13}$,  
so that it is natural to expect a smaller deviation as observed.
From eqs. (\ref{UeUnBm}) we obtain the following explicit expressions
for the observable quantities:
\bea
\tan^2 \theta_{23} &=& 1 + 4 s^e_{23} \cos(\alpha_2)\\
\delta_{sol} \equiv
1-\tan^2 \theta_{12} &=& 2 \sqrt{2} ( s^e_{12} \cos(\alpha_1) + s^e_{13} \cos(\delta_e+\alpha_2+\alpha_1))
\label{deltasol}\\
|U_{e3}| &=&\frac{1}{\sqrt{2}} ( {s^e_{12}}^2 +{s^e_{13}}^2 -2 \cos(\delta_e+\alpha_2) s^e s^e_{13} )^{1/2} 
\label{ue3}\\
\tan \delta &=& \frac{s^e_{12} \sin(\alpha_1) - s^e_{13} \sin(\delta_e+\alpha_2+\alpha_1) }
{s^e_{12} \cos(\alpha_1) - s^e_{13} \cos(\delta_e+\alpha_2+\alpha_1)}~~~~~,
\label{delta}
\eea
to be compared with the experimental data. 
According to \cite{Valle}
the 3-$\sigma$ windows are $|U_{e3}| \le 0.23$ and $0.36 \le \delta_{sol} \le 0.70$.

Notice that the sign of $\delta_{sol}$ is not necessarily positive, 
so that only a part (say half) of the parameter space in principle allowed for the phases 
is selected.
With the correction to $\delta_{sol}$ going in the good direction, 
one roughly expects $|U_{e3}| \sim \delta_{sol}/4 \approx 0.1-0.2$.
Hence, at present it is not excluded that charged lepton mixing can transform a 
bimixing configuration into a realistic one but there are constraints and, 
in order to minimize the impact of those constraints, 
$|U_{e3}|$ must be within a factor of 2 from its present upper limit. 
On the other hand, an upper limit on $|U_{e3}|$ smaller than $\delta_{sol}/4$
would start requiring a fine-tuning. 
Indeed, in order to reduce $|U_{e3}|$ significantly below $0.1-0.2$
a cancellation must be at work in eq. (\ref{ue3}), 
namely $\delta_e+\alpha_2$ should be close to $0$ or $2 \pi$ 
and $s^e_{12}$ and $s^e_{13}$ should be of comparable magnitude.
In addition, to end up with the largest possible $\delta_{sol}/4$, 
eq. (\ref{deltasol}) would also suggest a small value for $\alpha_1$.

\begin{figure}[!h]
\vskip .1 cm
\centerline{~~   
\psfig{file=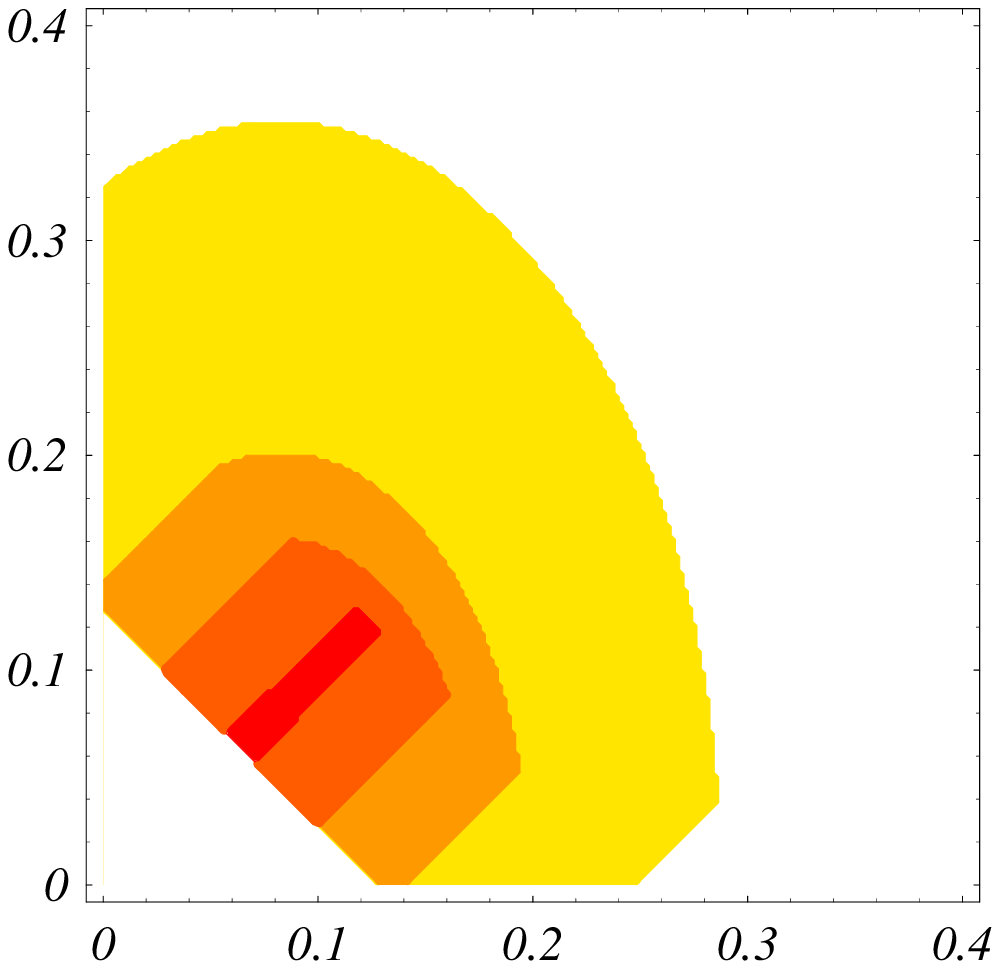,width=0.34 \textwidth}~~
\psfig{file=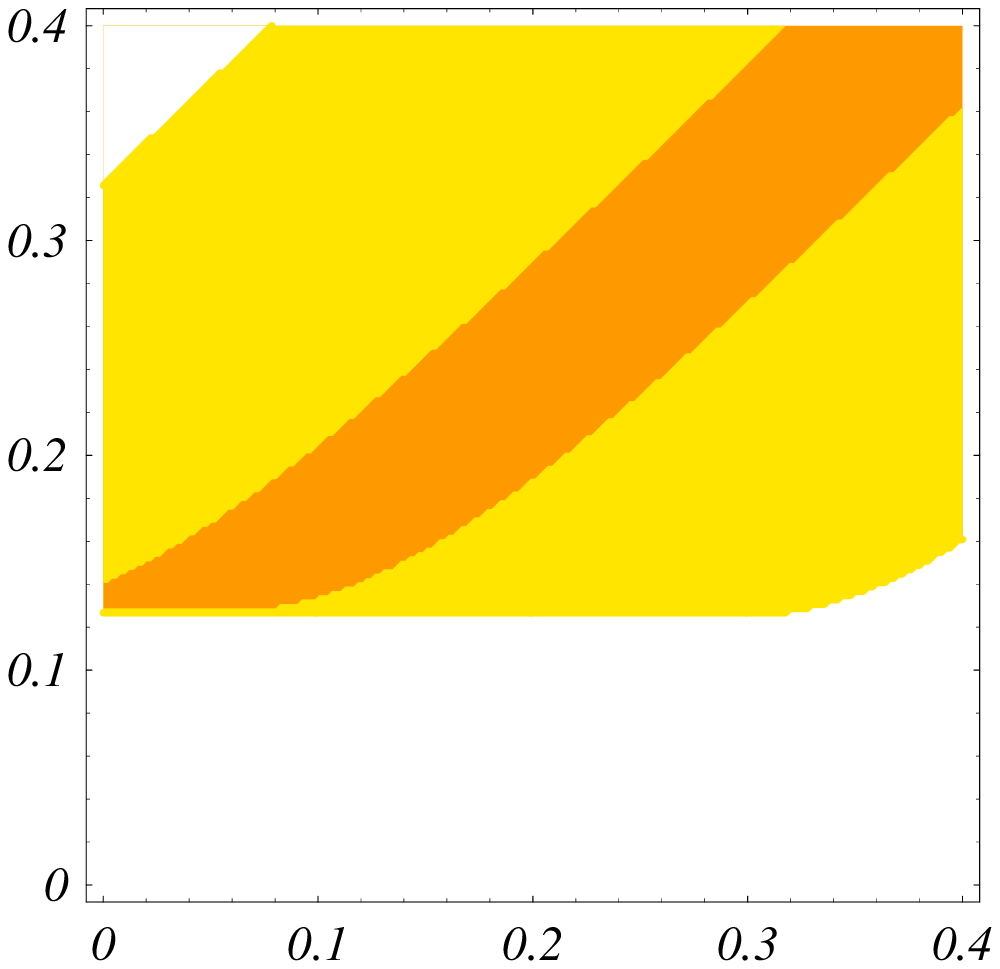,width=0.34 \textwidth}~~
\psfig{file=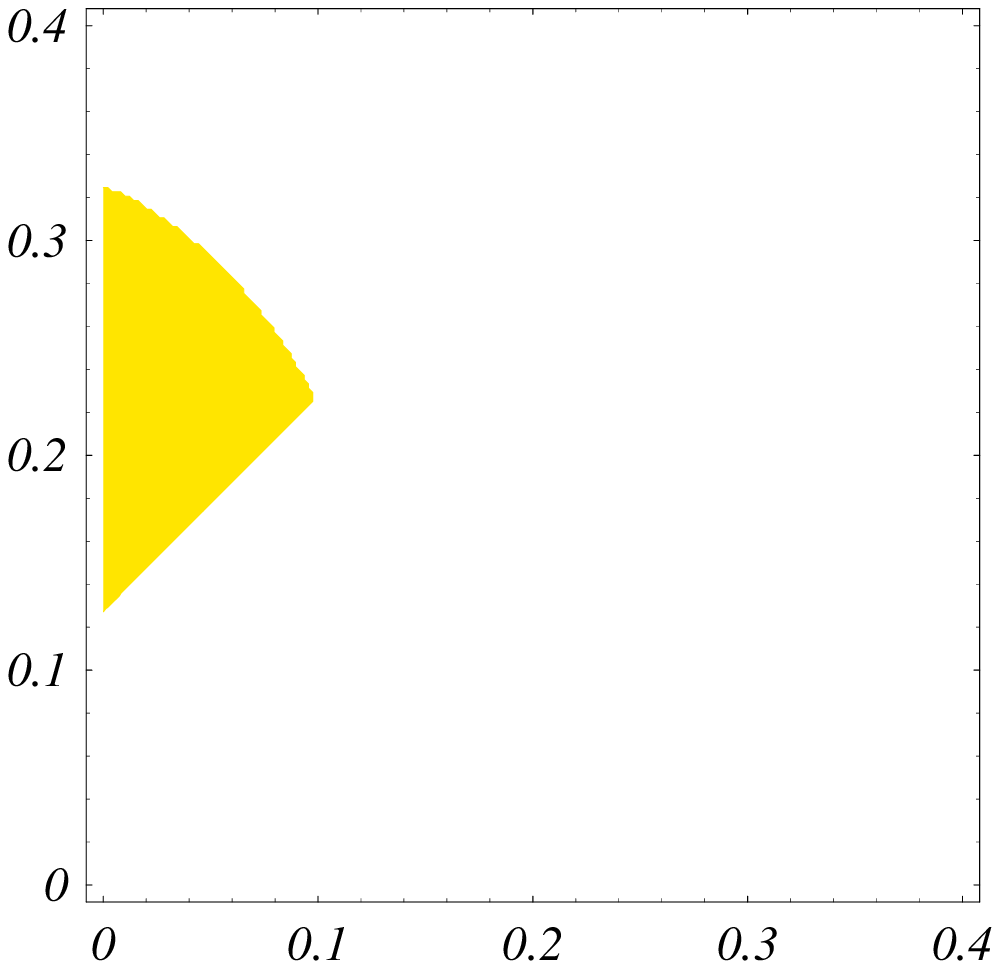,width=0.34 \textwidth} 
\put(-400, 150){ $\alpha_1=0$~~~~~~~~~~~~~~~~~~~~~ 
$\alpha_1=\pi /2$ (or $3 \pi/2$)~~~~~~~~~~~~~~~~~~~~~$\alpha_1=\pi$} 
\put(-360, -10){ $s^e_{12}$ } \put(-200, -10){ $s^e_{12}$ } \put(-50, -10){ $s^e_{12}$ } 
\put(-470, 120){ $s^e_{13}$ } \put(-313, 120){ $s^e_{13}$ }  \put(-156, 120){ $s^e_{13}$ }  
\put(-400, 110){ \footnotesize $.23$ }
\put(-190, 50){\footnotesize $.23$ }
\put(-125, 110){\footnotesize $.23$ }
\put(-420, 75){ \footnotesize $.1$ }
\put(-200, 95){\footnotesize $.1$ }
\put(-425, 60){ \footnotesize $.05$ }
\put(-420, 40){ \footnotesize $.01$ }
}
\caption[]{Taking an upper bound on $|U_{e3}|$ respectively equal to ${0.23, 0.1, 0.05, 0.01}$,
we show (from yellow to red) the allowed regions of the plane $[s^e_{12}, s^e_{13}]$.
Each plot is obtained by setting $\alpha_1$ to a particular value, 
while leaving $\alpha_2 + \delta_e$ free.
We keep the present 3 $\sigma$ window for $\delta_{sol}$ \cite{Valle}. }
\label{isole}
\end{figure}

The above considerations can be made quantitative by showing, 
for different upper bounds on $|U_{e3}|$,
the points of the plane $[s^e_{12}, s^e_{13}]$ which are compatible with the present
3 $\sigma$ window for the solar angle. 
This is shown in fig. \ref{isole}, where the three plots correspond to different choices
for $\alpha_1$. 
A point in the plane $[s^e_{12}, s^e_{13}]$ is excluded if there is no value 
of $\alpha_2 + \delta_e$ for which (\ref{ue3}) and (\ref{deltasol}) agree with experiment.
Regions in white are those excluded by the present bound on $|U_{e3}|$. 
With increasingly stronger bounds on $|U_{e3}|$, 
the allowed regions, indicated in the plots 
with increasingly darkness, get considerably shrinked.
For $|U_{e3}| \le 0.05$ only $|\alpha_1| < \pi/2$ is allowed.
Notice also that at present the two most natural possibilities $s^e_{12} \gg s^e_{13}$ and
$s^e_{12} \ll s^e_{13}$ are allowed but, with $|U_{e3}| < 0.1$, they are significantly constrained
and with $|U_{e3}| \le 0.05$ ruled out completely. 
Below the latter value for $|U_{e3}|$, a high level of degeneracy between 
$s^e_{12}$ and $s^e_{13}$ together with a small value for 
$\alpha_1$ and $\delta_e + \alpha_2$ are required. 

Summarising, planned improvements in the sensitivity to
$|U_{e3}|$ - which could reach the 0.05 level,
could have a crucial impact on bimixing models.
They could either disfavour it as unnatural (in the sense that a dynamical
principle or a symmetry acting also on the charged lepton mass matrix would 
have to be invoked) 
or, if $|U_{e3}|$ were to be found, support bimixing models.




%
\end{document}